\documentclass[journal,twocolumn]{IEEEtran}
\usepackage{graphicx}
\usepackage{amsmath}
\usepackage{amssymb }
\usepackage{graphicx}
\usepackage{amsmath}
\usepackage{amssymb}
\usepackage{amsthm}
\usepackage{lipsum}
\usepackage{cite}
\usepackage{amsmath,amssymb,amsfonts}
\usepackage{algorithmic}
\usepackage{graphicx}
\usepackage{textcomp}
\usepackage{xcolor}
\usepackage{array} 
\usepackage{soul,xcolor}
\setstcolor{red}
\usepackage[ruled,vlined]{algorithm2e}
\def\BibTeX{{\rm B\kern-.05em{\sc i\kern-.025em b}\kern-.08em
    T\kern-.1667em\lower.7ex\hbox{E}\kern-.125emX}}

\theoremstyle{definition}

\begin{document}

\title{Smart Interference Management xApp \\ using Deep Reinforcement Learning}
\author{Mahdi Eskandari,
        Shipra Kapoor,
        Keith Briggs,
        Arman Shojaeifard,~\IEEEmembership{Senior Member,~IEEE,} \\
        Huiling Zhu,~\IEEEmembership{Senior Member,~IEEE,}
        and~Alain Mourad
\thanks{M. Eskandari and H. Zhu are with the School of Engineering, University of Kent, CT2 7NT Canterbury, U.K. (e-mail:
me377@kent.ac.uk; H.Zhu@kent.ac.uk).}
\thanks{S. Kapoor and K. Briggs are with the BT Labs, Adastral Park, Ipswich IP5 3RE, U.K.
(e-mail: shipra.kapoor@bt.com; keith.briggs@bt.com)}
\thanks{A. Shojaeifard and A. Mourad are with the Interdigital, London EC2A 3QR, U.K. (e-mail: arman.shojaeifard@interdigital.com; alain.mourad@interdigital.com)}
}


\maketitle

\begin{abstract}
Interference continues to be a key limiting factor in cellular radio access network (RAN) deployments. Effective, data-driven, self-adapting radio resource management (RRM) solutions are essential for tackling interference, and thus achieving the desired performance levels particularly at the cell-edge. In future network architecture, RAN intelligent controller  (RIC) running with near-real-time  applications, called  xApps, is considered as a potential component to enable RRM.  In this paper, based on deep reinforcement learning (RL) xApp, a joint sub-band masking and power management is proposed for smart interference management. 
The sub-band resource masking problem is formulated as a Markov Decision Process (MDP) that can be solved employing deep RL to approximate the policy functions as well as to avoid extremely high computational and storage costs of conventional tabular-based approaches. The developed xApp is scalable in both storage and computation. Simulation results demonstrate advantages of the proposed approach over decentralized baselines in terms of the trade-off between cell-centre and cell-edge user rates, energy efficiency and computational efficiency.
\end{abstract}

\section{Introduction}
With the rapid growth in demand for high bandwidth applications in dynamic radio access networks (RAN), the need for improved spectrum efficiency has become indispensable, thus, triggering the need to develop new radio resource management (RRM) techniques. The (Open-RAN) O-RAN architecture is proposed to provide open platform to control operators to run cellular networks which could be built on network entities provided by multiple different vendors.
There are two fundamental pillars to O-RAN: Openness and Intelligence. Operators need open interfaces to introduce new services more quickly and to customize the network to meet their specific needs. Openness also facilitates multi-vendor deployments, enhancing competition in the industry. The open-source nature of software and hardware design facilitates faster, more efficient, and broader science and technology innovation while preserving backwards compatibility systems. 
 As wireless networks densify and applications become richer and more demanding, future wireless systems, including 5G and beyond 5G, will also become more complicated. As such, the vendors and the operators of mobile networks should self-organise. In order to automate network functions and reduce operational costs, they should utilize new technologies, such as Machine Learning (ML) and Artificial Intelligence (AI) \cite{gavrilovska2020cloud}.
  In O-RAN, flexibility, service orientation, and software-defined networking are key features. Additionally, artificial intelligence is a key component. The O-RAN architecture consists of several subsystems, as shown in Fig~\ref{sys_model}. Non-real time functionality is separated from real-time functionality, including service and model training for non-real time functionality \cite{singh2020evolution}. Additionally, real-time control functions are integrated into the RAN intelligent controller (RIC) for runtime execution of trained models and real-time control functions. Also, this layer is responsible for operations like interference management, quality-of-service (QoS) management, etc. 
Extensive studies on RRM have been reported in the literature, such as techniques
for sub-band and power allocation across different users \cite{MASKOOKI2015601, CWong1999, HZhu2009, chen2017machine}. That is, each base station (BS) can choose the users to be 
allocated to each sub-band, and can modify transmit power in orthogonal frequency-division multiplexing (OFDM) systems. Furthermore, to achieve an optimal network
performance, it is generally assumed that the channel state
information (CSI) is known by APs. However, in practical systems,
there are typically limited functionalities for inter-cell RRM such as (a) control of the transmit
power per cell, not per sub-band, (b) sub-band allocation per user not being possible, only the ability to mask (or turn the transmission on) individual sub-bands at individual cells, and (c) having reference signal strength measurements such as Reference Signal Received Power (RSRP), rather than detailed CSI. Therefore, with limited control over radio resource allocation and accessibility of radio parameters, conventional RRM techniques are not applicable to managing cellular networks in practice.

Comprehensive literature related to the usage of machine learning (ML) techniques can be found in \cite{chen2017machine}. Furthermore, deep reinforcement learning (Deep RL) has been used as function approximator to estimate the probability of taking each action given action space or value of each state-action pair. There is a rich literature related to deep RL algorithms application particularly in various gaming environments such as Atari and Go \cite{maddison2014move}. Deep RL also becomes a technology to solve problems in wireless communication \cite{ chen2019artificial,chiumento2016impact}. 

This paper proposes a framework of joint sub-band masking and power management to manage interference among multiple cells. The joint sub-band masking and power management is formulated as an optimisation problem to achieve the minimum power consumption while satisfying users’ data rate requirements. 
In the presented work, we have employed deep~RL to train the agent to take actions whilst interacting with the environment such that the reward over the time is maximised. 
The proposed deep~RL agent autonomously learns an optimal subset of resources to be masked based on the users’ channel condition such that the power consumption is significantly minimised whilst delivering a guaranteed QoS to all users. Next, each BS has ability to adjusts its total transmit power over all the sub-bands with respect to user location in the cell. The simulation results demonstrate that the proposed solution converges quickly and reduces the transmit power efficiently.

The contribution of the paper is summarised as follows
\begin{itemize}
    \item We formulate the optimisation problem as a power minimisation problem with the constraints of maximum power budget and minimum rate required for each user. The proposed  objective function is NP-hard duo to existence of integer parameter, for addressing this issue, the problem has been solved in two stages. first the sub-band masking problem is solved using the proposed deep RL algorithm, and then using the found sub-band allocations that is the output of deep RL model, the power minimisation is done in the second stage.
    \item In order to demosntrate the effectiveness of the proposed deep RL method as well as its robustness against different environment setups, in the simulations, we consider three different scenarios, (1) The case when the users in all of the cells are in the edge area of the which is the worst case that is the interference from the neighboring cell is maximum. (2) The case in which only users in one of the cells are on the edge. (3) The situation where no users are at the edges of the cells. In all of the considered scenarios, the proposed deep RL framework managed to allocate the sub-bands to the user with the target of keeping the rates of the users higher than the threshold.  
\end{itemize}
 
The paper is organised as follows: Section~II and section~III present the system model and discusses the problem formulation, respectively. Section~IV models the interference control process as an deep RL problem followed by power optimisation. Section~V describes the simulation environment and provides numerical results of the proposed scheme. Finally, section~VI concludes the paper.
\section{System Model}

\begin{figure}[t]
    \centering
    \includegraphics[scale=0.55]{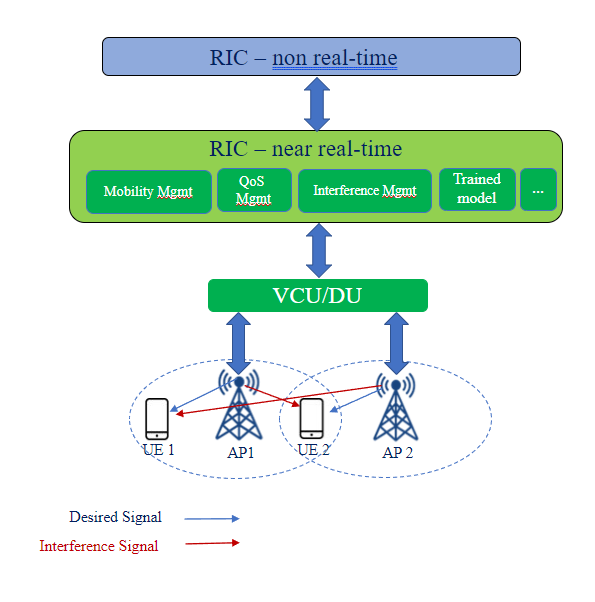}
    \caption{Representation of O-RAN architecture, with RAN intelligent
controllers (near real-time and non real-time).}
    \label{sys_model}
\end{figure}

We considere a downlink multi-cell OFDM network comprising $K$ cells and $U$ user equipments (UEs). The network architecture includes RIC component to enable RRM, as shown in Fig.~\ref{sys_model}. Henceforth, each cell is covered by one access point (AP) and managed by the RIC through the layer of distribute virtualised network functions (VNF). In each cell, a BS is located in the centre of the cell while UEs are either at cell centre or cell edge. 

Here, each AP intends to transmit data to its associated UEs using the $N$ masked sub-bands. User association is performed using max-RSRP method, that is, a UE will be associated with an AP which has the largest RSRP.  UEs receive a desired signal from their serving APs while experiencing interference from other APs. At any particular instant of time, all of the sub-bands could be exploited by each AP to serve its associated UEs. This implies that if the UE is at cell centre, then the impact of interference is not significant; therefore, all available sub-bands could be shared by neighbouring APs \cite{HZhu2014}. On the contrary, UEs particularly at the cell edge will experience sharp interference from neighbouring APs. To mitigate strong interference experiences by the UEs at the edge area, a set of sub-bands will be activated (masked) at each AP whilst  the rest of the sub-bands will be unmasked (deactivated). The mask indicator, $\beta _{k,n}$, is introduced to indicate the status of the sub-band $n$ in the cell $k$. If $\beta_{k,n} = 1$, sub-band $n$ is masked for all the UEs  in the cell $k$. Assuming partial information exchange between neighbouring APs, the channel quality indicator (CQI) measured by UE$_{k}$ under interference from AP$_{u}$, denoted by $ q_{u \rightarrow k,n} $, is known to both AP$_{k}$ and AP$_{u}$ employing which data rate is obtained using equation~(\ref{datarate_beta})

 \begin{equation}
                R_{k, u, n} =
        \left\{
        	\begin{array}{ll}
        		R_{k,n}(q_{k,n})  & \mbox{if } \beta_{u,n} = 0 \\
        		\hat{R}_{k,n}(q_{u\rightarrow k,n}) & \mbox{if } \beta_{u,n} = 1,
        	\end{array}
        \right .
        \label{datarate_beta}
\end{equation}
where $q_{k,n}$ is the CQI for the $u$-th UE at the $k$-th cell without interference.
Next, the achievable data rate for UE$_{u}$ in the cell $k$ over all the sub-bands is given by

\begin{equation}
    R_{k,u} = \sum_{n = 1}^{N} \beta_{k,n} \cdot R_{k, u, n}.
        \label{r_k}
\end{equation}
Thus, the total data rate of the network is 
\begin{equation}
    R = \sum_{k=1}^{K}  \sum_{u = 1}^{U} \sum_{n=1}^{N} \beta_{k,n} \cdot R_{k, u, n}.
    \label{r_all}    
\end{equation}
It can be concluded from (\ref{r_k}) and (\ref{r_all}) that user data rate will be changed if mask indicator is altered. Moreover, each UE periodically measure CQI, $q_{k,n}$, on sub-band $n$ and feeds back to the associated AP. The data rate achieved $\hat{R}_{k,n} (q_{k,n})$, from AP $k$ on sub-band $n$ corresponding to each CQI level $q_{k,n}$, can be obtained after selecting appropriate modulation and coding scheme (MCS) and signal-to-noise-ratio (SNR). Table~\ref{table:cqi} demonstrates MCS and SNR corresponding to 4-bit CQI whilst corresponding block error rate (BLER) is 10\%. It emerges clearly from Table~\ref{table:cqi}, that the CQI level can be improved or degraded by one level if transmit power is adjusted by about $2$ dB, thus making an impact on efficient data rate by 10\%.


\begin{table}[t]
\caption{CQI Table \cite{3GPPTS38}.} 
\centering 
\begin{tabular}{r c c r} 
\hline\hline 
SNR (dB) & CQI Index & Modulation & Code Rate ($\times 1024$) \\ [0.5ex] 
\hline 
$-\infty$ & $0$ & Out of Range & $-$\\
$-6.9360$ & $1$ & QPSK & $78$  \\
$-5.1470$ & $2$ & QPSK & $120$ \\
$-3.1800$ & $3$ & QPSK & $193$ \\
$-1.2530$ & $4$ & QPSK & $308$ \\
$0.7610$ & $5$ & QPSK & $449$ \\
$2.6990$ & $6$ & QPSK & $602$ \\
$4.6940$ & $7$ & 16-QAM & $378$ \\
$6.5250$ & $8$ & 16-QAM & $490$ \\
$8.5730$ & $9$ & 16-QAM & $616$ \\
$10.3660$ & $10$ & 64-QAM & $466$ \\
$12.2890$ & $11$ & 64-QAM & $567$ \\
$14.1730$ & $12$ & 64-QAM & $666$ \\
$15.8880$ & $13$ & 64-QAM & $772$ \\
$17.8140$ & $14$ & 64-QAM & $873$ \\
$19.8290$ & $15$ & 64-QAM & $948$
 \\[1ex]  
\hline  
\end{tabular}
\label{table:cqi} 
\end{table}
The transmit power of BS $k$ is denoted by $p_k$, limited to a maximum value of $p_{\mathrm{max}}$, i.e., $p_k \leqslant p_{\mathrm{max}}$, also assuming the transmit power for channel measurement is $p_{\mathrm{max}}/N$ for each sub-band. Furthermore, considering practical power management constraint, the transmit power is calibrated per BS rather than per sub-band. 
Next, after power tuning, the CQI on each sub-band is monitored again and fed back to the BS to estimate the CQI for corresponding sub-band, as well as the achievable data rate. 

\section{Problem Formulation}
Our goal is to propose an algorithm that outputs a set of sub-bands to be masked at each BS across the network such that power consumption is significantly minimised whilst a guaranteed QoS is delivered to all the UEs. This is achieved by satisfying a data rate constraint $R_\mathrm{min}$, which is assumed to be the same for all the UEs. As in the network, each UE is exclusively served by its serving BS, the data rate of each user will be determined by the transmit power of the serving BS and interference from neighbouring APs. Let $\mathcal{K}$ denote the set of cells where $\mathcal{K}=\{0, 1, \dots, K-1\}$, $\mathcal{N}$ denote the set of sub-bands where $\mathcal{N} = \{0, 1, \dots, N-1\}$ and $\mathcal{U} = \{0, 1, \dots, U-1\}$ as the set of users. It is evident from (\ref{r_k}) that for UE$_{k}$, data rate can be changed by varying mask indicator of sub-bands. Thus, with the objective of using minimum transmit power to satisfy data rate constraint for all users, the joint sub-band masking and power allocation problem in a multi-cell network as optimisation problem is formulated as
\begin{align}
\label{opt}
    \min_{ \{ \beta_{k, n}, p_k\}, k \in \mathcal{K}, n \in \mathcal{N}}  p_k, \\ \nonumber
     \mathrm{s.t.} \hspace{3mm}  p_k \leqslant p_{\mathrm{max}},  \text{ for all }k\in \mathcal{K} &\\ \nonumber
     R_{k,u} \geqslant R_{\mathrm{min}}, \text{ for all }k\in \mathcal{K} \text{ and }u\in \mathcal{U}&.
\end{align}
Solving (\ref{opt}) is NP-hard as it includes a integer parameter, $\beta_{k,n}$. If there are $M$ main interference sources (APs) for a reference edge user, the complexity of exhaustive searching will be $\mathcal{O}((3M)^N)$, i.e., each interfering BS has three choices for each sub-band (a) mask (b) unmask and (c) share the sub-band with reference UE. In the developed algorithm, the joint optimisation problem is solved using two stage iteration. In the first iteration sub-band masking is performed  followed by power allocation optimisation in second iteration.

\section{Solution based on Deep RL}
\subsection{Deep RL Background}
Deep RL is used to solve sub-band allocation problem where the agent learns a set of optimal sub-bands to be masked at each BS. At each time step, whenever CQI are updated at the APs, a new episode $j$ starts. $T$ is defined as the number of time steps in each episode. At each time step $t$, the agent interacts with the environment $\mathcal{E}$, selects an action $a_t$ from valid choices of actions, receives corresponding reward $r_t$ and observes new state $s_{t+1}$. The goal of the RL agent is to maximise discounted future rewards. The action-value function under a given policy $\pi$ is expressed as $Q_\pi(s, a)$. It is the expected discounted reward at the start when in state $s$ to learn an action $a$ under the policy $\pi$. The optimal action-value function is defined as $Q ^* = \max_\pi \mathbb{E}[R_t | s_t = s, a_t = a, \pi]$, where $R_t = \sum_{t^\prime = t} ^{T} \gamma^{t^\prime-t}r_{t^\prime}$, $T$ is the time-step until the termination of the episode, and $\gamma$ is the discount factor to the reward. The optimal action-value function $Q^*(s, a)$ obeys the identity called the Bellman equation, given by
\begin{equation}
    Q^*(s, a) = \mathbb{E}_{s^\prime \sim \mathcal{E}} [r + \gamma \max_{a^\prime}Q^*(s^\prime, a^\prime) | s, a].
\end{equation}
A key approach adopted here is an iterative procedure for finding an ideal estimate of action-value function, i.e., $Q_{i+1}(s, a) = \mathbb{E} [r + \gamma \max_{a^\prime}Q_i(s^\prime, a^\prime) | s, a]$, where $i$ is the index of iteration for updating the value of $Q$. With continuous policy evaluation and policy update iterations, an optimal policy is learnt, i.e., $Q_i \rightarrow Q^*$ as $i \rightarrow \infty$ \cite{sutton1998introduction}. In practice, neural networks with weights denoted by $\boldsymbol{\theta}$ could be used as a function approximator to estimate the action-value function \cite{mnih2013playing}. At each iteration $i$, $Q_i(s, a)$ is represented by $Q(s, a; \boldsymbol{\theta}_i)$, and the network is trained by minimising the loss function $\mathcal{L}_i(\boldsymbol{\theta}_i)$ defined as
\begin{equation}
    \mathcal{L}_i(\boldsymbol{\theta}_i) = \mathbb{E}_{s, a \sim \varrho (.)}[(y_i - Q(s, a; \boldsymbol{\theta}_i))^2],
    \label{eq_loss}
\end{equation}
where $y_i = \mathbb{E}_{s^\prime \sim \mathbb{E}} [r + \gamma \max_{a^\prime}Q(s^\prime, a^\prime; \boldsymbol{\theta}_{i-1}) | s, a]$ is the target for the $i$-th iteration and $\varrho (s, a)$ is a probability distribution over $s$ and $a$.

This algorithm is \emph{model-free} because it does not construct an estimate of the environment $\mathcal{E}$; furthermore, it utilizes an exploration-based behaviour to discover greedy strategies, causing it to be an \emph{off-policy} algorithm.

\subsection{Sub-band masking using deep RL}
Generally, deep RL is presented as a Markov decision process (MDP) with observation and action spaces. When solving the sub-band masking problem, this cellular system is denoted by the environment $\mathcal{E}$, while agent is RIC which is able to control all the cells. The following are the key RL elements that are employed to solve sub-band allocation problems.
\subsubsection{Observation space}
    At each time step $t$, the observation consists of two types of components, the sum-rates of all the users of all of the cells over all of the sub-bands and a vector containing the CQI levels of all the UEs at of the cells. Note that when a sub-band is unmasked, the CQI level of that particular sub-band will set to be zero. Hence, the observation vector will be expressed as
    \begin{equation}
        \mathbf{s}_t = \{\mathcal{R}, \mathcal{Q} \},
    \end{equation}
    where $\mathcal{R} = [{R_{1,1}, \dots, R_{1,U}, R_{2,1}, \dots, R_{2,U}, \dots, R_{K,U}} ]$ is the vector of sum-rates on all $N$ sub-bands for all the users. $\mathcal{Q} =[\beta_{1,1},\dots,\beta_{1,N}, \dots. \beta_{K,N}]$ is a binary vector indicating the sub-band masking indicator for each cell on all the sub-bands. Finally, the observation shape is $K(U+N)$.
    \subsubsection{Action space}
    The action space consists of masking and unmasking a sub-band for each cell. The action space is $KN+1$, where the constant part represents an action that the agent keeps the previous CQI arrangements and does not make any changes in masking or unmasking a sub-band. The rest $KN$ actions are the index of a particular sub-band of each BS that is going to be masked or unmasked. The index of the target cell is $\lfloor \frac{a_t}{N} \rfloor$ where $a_t$ is the action taken by the agent. Furthermore, the target sub-band for performing action is $\mod{(a_t, KN)}$. The execution of each action on the user is by toggling the mask indicator, as a result, when a sub-band is masked, by execution of action in will unmasked and vice versa. 
     \subsubsection{Reward function}
     At every time step $t$, the reward is calculated based on the difference between current rate for user $u$ at cell $k$ and minimum rate required by each UE by following $\Delta R_{k, u}^{(t)} = {R}_{k,u} - R_{\mathrm{min}}$, where $R_{\mathrm{min}}$ is the minimum rate required for each UE. The reward for each UE $u$ at cell $k$ can be calculated as
    \begin{equation}
                r_{u,k}^{(t)} =
        \left\{
        	\begin{array}{ll}
        		0  & \mbox{if } \Delta R_{u,k}^{(t)} \geqslant 0 \\
        		\Delta R_{u,k}^{(t)} & \mbox{if } \Delta R_{u,k}^{(t)} < 0,
        	\end{array}
        \right.
    \end{equation}
hence, the reward function at timestep $t$ is given by
            \begin{equation}
               \Bar{r}_t = \sum_{k=1}^{K} \sum_{u=1}^{U} {r_{u,k}^{(t)}}.
            \end{equation}
The reward is always negative for the agent since it is the difference between the current rate and the minimum rate and the agent should try to minimise the negative reward. When the rate touches the minimum rate, the new reward would be zero.
Finally, an additional reward of $\rho$ is given to the agent when the following conditions hold simultaneously.  \textit{First, The current action was to do nothing. Second, when $\Bar{r}_t = 0$}. The agent's goal is to increase the rate of all the UEs in all the cells above the minimum rate required. In general, multiple CQI arrangements are possible that satisfy the requirements of minimum rates for all UEs, so if the agent finds one of those arrangements, that would suffice. Adding this additional reward encourages the agent to remain at the stable status that it has reached and avoids fluctuating towards other conditions.   
The details of the proposed deep Q-learning algorithm for sub-band masking are presented in Algorithm~\ref{alg_1}.

\begin{algorithm}[h]
\caption{Deep RL for sub-band masking.}
\label{alg_1}
\SetAlgoLined
 \textbf{Initialisation:} Initialise  time, states, actions, and replay buffer $\mathcal{D}$ for storing the tandem states, action and reward in each time step\;
 \textbf{Output:} Allocation of sub-bands between users \;
 \For{$\mathrm{episode}\hspace{1mm} j = 1, \dots, J$}{
  Initialise the environment $\mathcal{E}$ and obtain the current CQIs for all the users on all the sub-bands and make the initial state $s_1$\;
  Set $t = 0$ \;
  \While{$t \leq T$}{
   $t = t +1$\;
   Observe current state $s_t$ \;
   $\epsilon := \max(\epsilon; \epsilon_\mathrm{min})$ \;
   Sample $\delta \sim \mathrm{Uniform}(0, 1)$ \;
   \eIf{$\delta < \epsilon$}{
   Select action $a_t$ randomly\;}
   {Select an action $a_t = \arg \max_{a^\prime}  Q (s_t, a^\prime; \boldsymbol{\theta}_t)$ \;} 
   Observe next state $s_{t+1}$ and reward $r_t$, Then, store transition set
$\{s_t, a_t, r_t, s_{t+1} \}$ into $\mathcal{D}$ \;
Sample random mini-batch of transitions $\{s_j, a_j, r_j, s_{j+1}\}$
from $\mathcal{D}$ \;
Set $y_j = r_j$ for terminal point and $y_j = r_j + \gamma \max_{a^\prime} Q(y_{j+1}, a^\prime; \boldsymbol{\theta}_t)$ for non-terminal point \;
Perform stochastic gradient descent on $(y_j - Q(s_j, a_j; \boldsymbol{\theta}_t))^2$ to find $\boldsymbol{\theta}^*$ \;
Update $\boldsymbol{\theta}_t := \boldsymbol{\theta}^*$  \;
Set $s_t := s_{t+1}$ \;
$\epsilon := \epsilon - \epsilon_d$ \;
   }
 }
 Save the deep RL model \;
\end{algorithm}

\subsection{Power management}
Power management is applied after the RL model is trained to mask a set of sub-bands at each BS. The aim of power management is to adjust the transmit power of all the APs to the lowest value such that a guaranteed data rate could be achieved by each UE. 
The power adjustment process is described in Algorithm~\ref{alg_2} in detail. To begin with, one of the cells with the highest rate of UEs is selected as the target AP, and the power reduction process is initiated there. At each stage, the power budget of the target cell is reduced by one step, and the rate of UEs is measured. The reduction of power in one cell will cause the rate of the UEs in that cell to decrease and the rate of UEs using that same sub-band in other cells to increase, since reducing power in one cell will also reduce interference in all other cells. This procedure continues until the rate of at least one of the users in the target cell touches the minimum rate. Then another AP with a similar condition is chosen as the target cell, and the same procedure is followed. It is important to note that decreasing the power of one cell could increase the rate of UEs in another cell which was earlier targeted. This means that an AP may be chosen as the target AP more than once. 
 \begin{algorithm}
\SetAlgoLined
  \textbf{Output:} Minimum power required $p_{\mathrm{opt}, c}$ and $p_{\mathrm{opt}, e}$ for the user in the interfering cell and reference user, respectively \;
 Load the trained DQN weights obtained from algorithm~\ref{alg_1} \;
 Initialise a new environment and obtain the current CQIs for all the users on all the sub-bands \;
 Apply the trained model on the environment and get the rates of the users after sub-band masking \;
 Set $p_{\mathrm{opt}, i} = p_\mathrm{max}$ $\forall{i}\in \mathcal{K}$\;
 Let $R_\mathrm{j,i}$ be the sum-rate of user $j$ in cell $i$ on all the sub-bands \;
 Set $\tilde{i} = \arg \max_iR_\mathrm{j,i}$\;
 \While{$\min_j  R_\mathrm{j,\tilde{i}}  > R_{\mathrm{min}}$ }{
  {
    Set $p_{\mathrm{opt}, \tilde{i}} = p_{\mathrm{opt}, \tilde{i}} - p_\mathrm{s}$ \;
    Obtain $R_\mathrm{j,i} \forall{i} \in \mathcal{K}$ and $j \in \mathcal{U}$ corresponding to new transmit powers \;
    Set $\tilde{i} = \arg \max_iR_\mathrm{j,i}$\;
   }
 }
 \caption{Power management.}
 \label{alg_2}
\end{algorithm}

\section{Simulation Results}

\begin{table}[ht]
\caption{Reinforcement Learning  Hyperparameters} 
\centering 
\begin{tabular}{l r } 
\hline\hline 
Parameter & Value \\ [0.5ex] 
\hline 
Discount factor $\gamma$ & $0.995$  \\
Number of first fully connected layer & $128$ \\
Initial exploration rate $\epsilon$ & $1.00$  \\
 Number of second fully connected layer & $128$ \\
Exploration rate decay $\epsilon_d$ & $0.000008$ \\
Learning rate & $0.0001$ \\
Minimum exploration rate $\epsilon_{\mathrm{min}}$ & $0.01$ \\
Extra reward $\sigma$ & $2$ \\
batch size & $32$ \\
Maximum time-steps of each episode $T$ & $64$ \\
Number of episodes & $5000$ \\
Activation function for hidden layers & ReLU \\
Activation function for output layer & Linear \\
Base station (AP) maximum transmit power $p_{\mathrm{max}}$ & $40$ dBm \\
 Antenna gains (AP, UE) & $(0, 0)$ dBi \\
Path loss exponent & $3$ \\
Power reduction step size $p_\mathrm{s}$ & $0.5$ dBm \\
Downlink frequency band & $2.8$ GHz \\
Number of sub-bands $N$ & 8 \\
Total bandwidth & $20$ MHz \\
Cell radius & $400$ m \\
Edge area & $20\%$ of cell radius \\
Minimum rate required $R_\mathrm{min}$  & $17.82$ Mbps   \\
Number of users per cell & $1$ \\
Noise power  & $-150$ dBm/Hz   \\
 [1ex]
\hline 
\end{tabular}
\label{table:rl} 
\end{table}

\begin{figure*}[!htb]
\minipage{0.45\textwidth}
  \includegraphics[width=\linewidth]{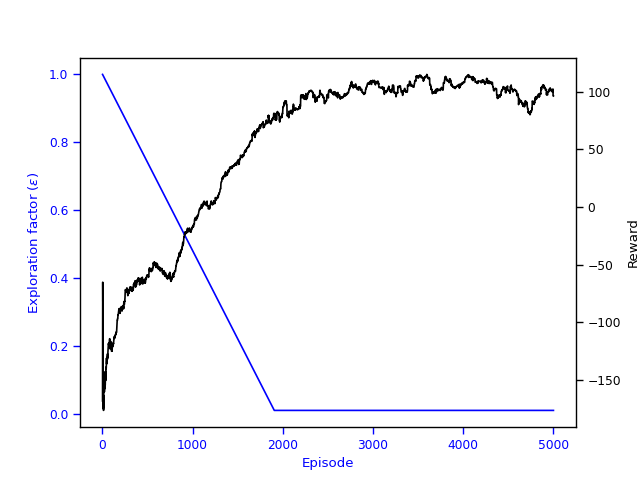}
  \caption{Rewards and exploration rate over episodes}\label{reward_one_edge}
\endminipage\hfill
\minipage{0.45\textwidth}
  \includegraphics[width=\linewidth]{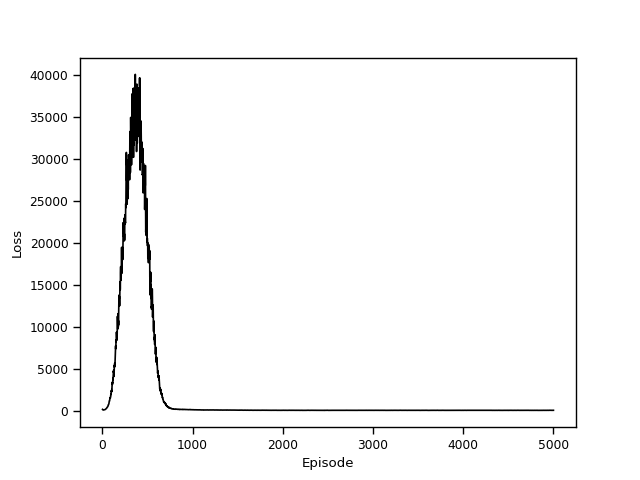}
  \caption{Total loss of the network as a function of number of episodes}\label{loss}
\endminipage\hfill
\minipage{0.32\textwidth}%
\endminipage
\end{figure*}

In this section, the performance of the proposed deep RL algorithm for joint optimisation of interference with power is assessed. The simulation environment is run on Python 3.8.9 with TensorFlow 2.4.0 on a computer with AMD Ryzen 7PRO Eight-Core Processor 3.20 GHz CPU and $32$ GB of memory. The hyperparameters for training the agent and the parameters for generating the system model are listed in Table~\ref{table:rl}. Each episode of the simulations resets the UE locations and as a result all the channel realisations.
Fig.~\ref{reward_one_edge} presents the cumulative reward and decay in exploration rate $\epsilon$ with respect to number of episodes. 
 To remove rapid fluctuations in output, a running average over $100$ steps is used. The exploration rate decays per episode, and subsequently approaches $0.01$ as the simulation runs. The average reward approaches a maximum value after running almost $2000$ episodes, indicating convergence. The proposed algorithm also demonstrates good adaptation to changes in channel conditions, as updates in CQIs at every episode lead to changes in environment. Fig.~\ref{loss} shows the loss function in Eq. ~(\ref{eq_loss}), from the graph, it can be seen that the loss at the beginning of the learning is almost zero, as all the actions are random and the network doesn't produce any action, but as the episodes progress, the loss increases as the network weights are updated and the number of random actions decreases. Over time, the loss begins to decrease as the agent learns more and more and the weights are becoming more accurate. At last, the loss becomes zero. 
 In the simulations, two neighbouring BS are considered, each serving one user. The simulation is done in three case studies, (1) all the UEs are in the edge area, (2) one of the UEs in the edge area and (3) all the UEs are in outside of the edge area.

\subsection{Case 1}

\begin{figure*}[!htb]
\minipage{0.32\textwidth}
  \includegraphics[width=\linewidth]{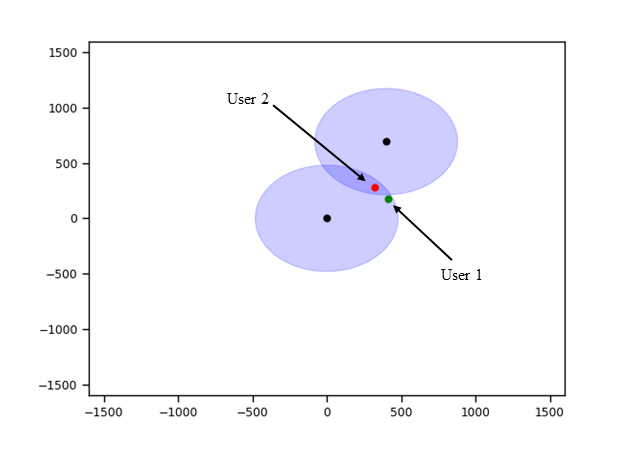}
  \caption{Cellular system illustration for case study 1 when one user is in the edge area and the other is in the centre area of the cell. A user in the overlapping area of two cells is coloured red and otherwise it is shown in green.}\label{case_1_env}
\endminipage\hfill
\minipage{0.32\textwidth}
  \includegraphics[width=\linewidth]{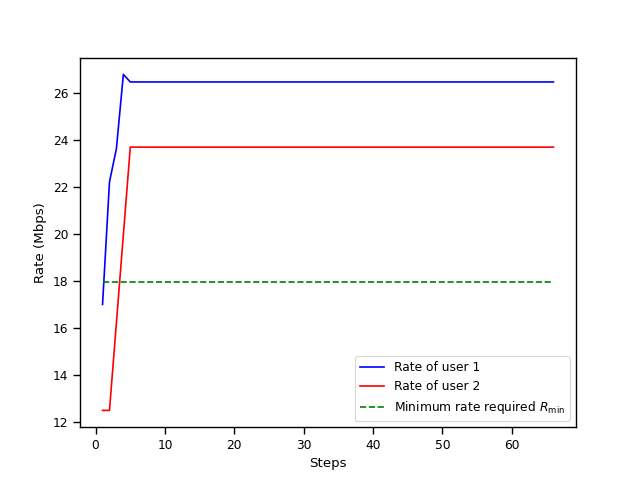}
  \caption{Illustration of the change in the rate of UEs when the learnt agent is applied to the cellular environment in Fig.~(\ref{case_1_env})}\label{rate_case1}
\endminipage\hfill
\minipage{0.32\textwidth}%
  \includegraphics[width=\linewidth]{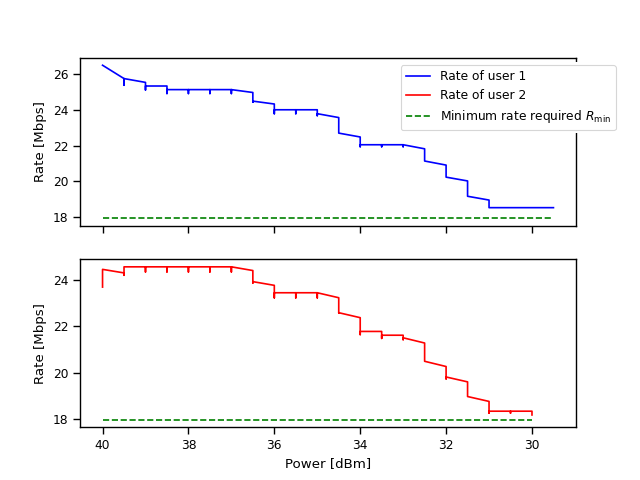}
  \caption{Power management for the users based on algorithm~\ref{alg_2}}\label{power_level2}
\endminipage
\end{figure*}

The users in this case as shown in Fig.~\ref{case_1_env}, are located near the overlapped areas of two cells, which result in severe interference from the neighbouring cell. Fig.~\ref{rate_case1} shows the change of the rates of all the users when the learnt agent is applying to the corresponding APs. The initial rate of all UEs is lower than the minimum threshold, but after the sub-allocation has been done, the rate of all UEs is higher than the minimum rate required. Additionally, the agent does nothing after the sub-band allocation period, i.e., after timestep 8. Next, as in algorithm \ref{alg_2}, the starting AP for power reduction is the cell that provides the highest rate, so the AP for user~1 is selected, then power reduction starts, and the rate of user~1 decreases, while the rate of user~2 increases. In this process, the power reduction algorithm continues until the rate of user~2 reaches the minimum rate required. 

\subsection{Case 2}
\begin{figure*}[!tb]
\minipage{0.32\textwidth}
  \includegraphics[width=\linewidth]{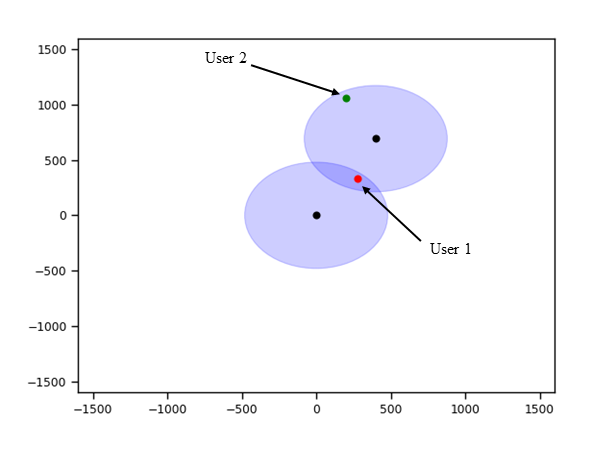}
  \caption{Illustration of the system model for case study 2 when one user is in the edge area and the other is in the centre area of the cell.}\label{case2_model}
\endminipage\hfill
\minipage{0.32\textwidth}
  \includegraphics[width=\linewidth]{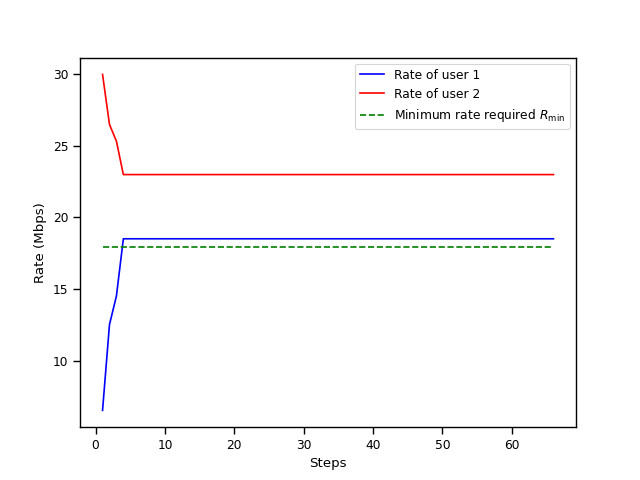}
  \caption{Applying the learned agent to the considered system model in order to modify the rate of UEs}\label{rate_case2}
\endminipage\hfill
\minipage{0.32\textwidth}%
  \includegraphics[width=\linewidth]{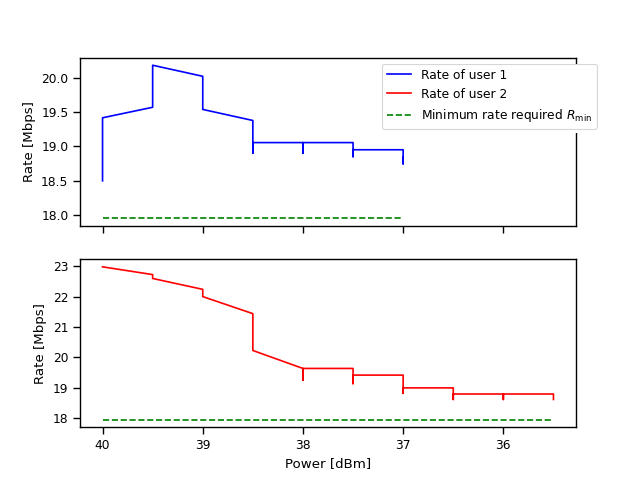}
  \caption{Power management for the users of case 2 based on algorithm~\ref{alg_2}}\label{power_level_case2}
\endminipage
\end{figure*}

In this case study, it is assumed that one of the users located in the centre area of the cell while the other one is in the edge of two cells and experiencing interference from other cell as shown in Fig. \ref{case2_model}. By applying the agent to the system, as illustrated in Fig.~\ref{rate_case2}, at the beginning, the rate of edge user, (user 1 in the graph), is lower than minimum rate and the rate of other user is higher than threshold, but, by applying the agent to the system, the rates of all the user fixed to a point higher than minimum rate which shows the accuracy of the agent. After allocating the sub-bands among the cells, in the power management part, the process begins with reducing the power of the cell that serves user 2, hence the rate of user 1 increases and then the power of AP 1 begins to decrease, this process goes on until the rate of user 2 touches the minimum rate. 

\subsection{Case 3}
\begin{figure*}

\minipage{0.32\textwidth}
  \includegraphics[width=\linewidth]{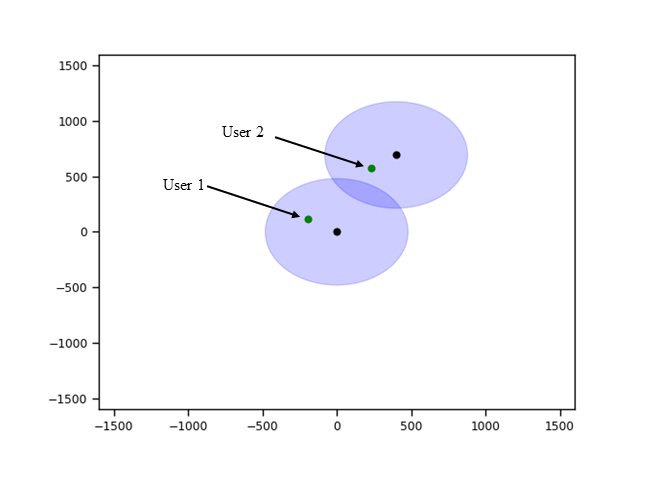}
  \caption{System model for case study 3 where both of the users are in centre area of the cells}\label{case3_model}
\endminipage\hfill
\minipage{0.32\textwidth}
  \includegraphics[width=\linewidth]{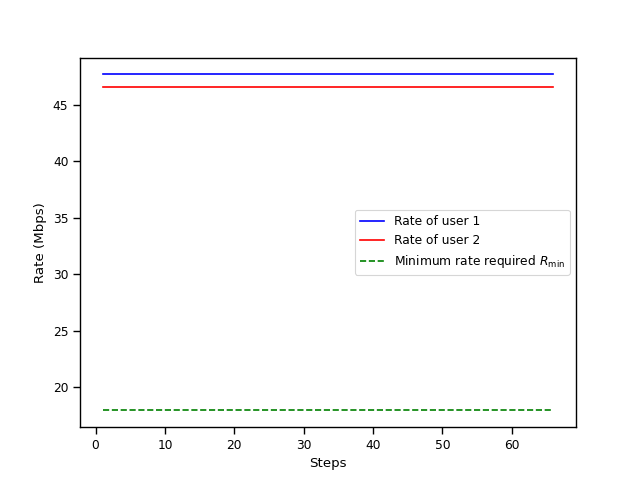}
  \caption{Testing the agent to the system model of case study 3}\label{rate_3}
\endminipage\hfill
\minipage{0.32\textwidth}%
  \includegraphics[width=\linewidth]{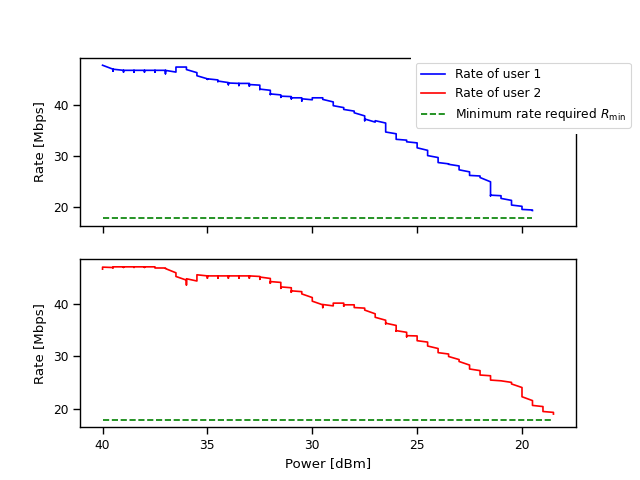}
  \caption{Power management for case 3 based on algorithm~\ref{alg_2}}
  \label{power_level3}
\endminipage
\end{figure*}

In this case, all of the users are located in the centre area of the cell where there is minimum interference from other cells, the system model is shown in Fig.~\ref{case3_model}. As shown in Fig,~\ref{rate_3}, initially, the rate of the users are higher than the minimum rate and the agent skip the sub-band allocation and does nothing which is reasonable based on the shaping of the rewards. Hence, all the rates remain unchanged during sub-band masking and all the sub-bands are shared among the users. Finally for power allocation part, again, the cell with the highest provided rate is picked for the power management and the algorithm~\ref{alg_2} is applied for power reduction. The power reduction process is shown in Fig.~\ref{power_level3}.

\section{Conclusion}
This paper presents a novel deep RL xApp for performing joint optimisation of radio resource management with power, considering the practical constraints of power management whilst delivering guaranteed QoS to all mobile users in cellular networks. It is formulated as an optimisation problem and solved using deep Q-learning. The following conclusions can be drawn from the simulation results using the proposed solution:
\begin{itemize}
\item When the distance between users in adjacent cells is relatively large, more sub-bands are shared.
\item Compared to a central user, an edge user requires more sub-bands to be deactivated at the interfering AP.
\item The power consumption is reduced significantly, while respecting the data rate constraints.
\end{itemize}

\bibliographystyle{IEEEtran}
\bibliography{refs}

\end{document}